# Nonlocal Equations for the Electromagnatic field in the Coupled Cavity Model

M.I. Ayzatsky[1]
National Science Center Kharkov Institute of Physics and Technology (NSC KIPT),
610108, Kharkov, Ukraine

The procedure for obtaining a difference equation, the solution of which is the components of the electric (or magnetic) field at the chosen set of volume points of the resonator chain, was developed. We started with the wave equation with boundary conditions and obtained the difference equation without boundary conditions. Boundary conditions were included into the coefficients of the difference equation for the electric field. As the obtained equation connects the field values in different points (in general case, on an infinite set of points), the proposed procedure represents a nonlocal model for field description. Solutions of the difference equation for the electric field were analyzed. It was shown that they coincide with good accuracy with the ones that were obtained by direct summing of relevant series.

## 1 Introduction

In the frame of the Coupled Cavity Model (CCM) the electromagnetic field in each cavity of the chain of resonators is represented as the expansion with the short-circuit resonant cavity modes [1,2,3,4,5,6,7,8,9,10,11,12,13]

$$\vec{E}^{(k)} = \sum_q e_q^{(k)} \vec{E}_q^{(k)}(\vec{r}) , \qquad (1)$$

where $e_q^{(k)}$ - amplitudes of the $q$ modes, $q = \{m, s, n\}$.

For a long time, the most used models were based on the assumptions that if $\omega \sim \omega_{q_0}^{(k)}$ ($\omega_{q_0}^{(k)}$ - eigen frequency of the short-circuit $k$-cavity mode) only one term with $q = q_0$ in the expansion (1) can be taken into account

$$\vec{E}^{(k)} \approx e_{q_0}^{(k)} \vec{E}_{q_0}^{(k)}(\vec{r}) , \qquad (2)$$

and the RF coupling of a $k$-cavity can be restricted to two ($k+1, k-1$) neighboring resonators

$$\left(\omega_{q_0}^{(k)2} - \omega^2\right) e_{q_0}^{(k)} = \omega_{q_0}^{(k)2} \alpha_{q_0,q_0}^{(k,k)} e_{q_0}^{(k)} + \omega_{q_0}^{(k)2} \alpha_{q_0,q_0}^{(k,k+1)} e_{q_0}^{(k+1)} + \omega_{q_0}^{(k)2} \alpha_{q_0,q_0}^{(k,k-1)} e_{q_0}^{(k-1)} . \qquad (3)$$

where $\omega_{q_0}^{(k)}$ - eigen frequencies of these modes.

Rejecting this assumption and considering all members of the expansion (1), such coupled equations for $e_{q_0}^{(k)}$ can be obtained under choosing the mode $q = q_0$ as the basic one [11,14,15,16]

$$Z_{q_0}^{(k)} e_{q_0}^{(k)} = \sum_{j=-\infty}^{\infty} e_{q_0}^{(j)} \alpha_{q_0,q_0}^{(k,j)} . \qquad (4)$$

Here $e_{q_0}^{(k)}$ - amplitudes of the $q_0$ modes, $Z_{q_0}^{(k)} = 1 - \frac{\omega^2}{\omega_{q_0}^{(k)2}}$, $\omega_{q_0}^{(k)}$ - eigen frequencies of these modes, $\alpha_{q_0,q_0}^{(k,j)}$ - real coefficients that depend on both the frequency $\omega$ and geometrical sizes of all resonators.

The CCM is a kind of decomposition approach in the study and development of complex electrodynamic systems [17,18,19]

Sums in the right side of (4) can be truncated

---

[1] M.I. Aizatskyi, N.I.Aizatsky; aizatsky@kipt.kharkov.ua

$$Z_{q_0}^{(k)} e_{q_0}^{(k)} = \sum_{j=k-N}^{k+N} e_{q_0}^{(j)} \alpha_{q_0,q_0}^{(k,j)}. \tag{5}$$

In the frame of the CCM the coupling coefficients $\alpha_{q_0,q_0}^{(k,j)}$ are electrodynamically strictly defined for arbitrary $N$ and can be calculated with necessary accuracy. In the theory of RF filters the coupling matrix circuit model is used intensively (see, for example, [20] and cited there literature). The main problem is how to calculate the matrix elements.

Amplitudes of other modes ($q \neq q_0$) can be found by summing the relevant series

$$e_q^{(k)} = \frac{\omega_q^{(k)2}}{\omega_q^{(k)2} - \omega^2} \sum_{j=-\infty}^{\infty} e_{q_0}^{(j)} \alpha_{q_0,q}^{(k,j)} \approx \frac{\omega_q^{(k)2}}{\omega_q^{(k)2} - \omega^2} \sum_{j=k-N}^{k+N} e_{q_0}^{(j)} \alpha_{q_0,q}^{(k,j)}. \tag{6}$$

Electric field in the resonator can be calculated as

$$\vec{E}^{(k)}(\vec{r}) = e_{q_0}^{(k)} \vec{E}_{q_0}^{(k)}(\vec{r}) + \sum_{q \neq q_0} e_q^{(k)} \vec{E}_q^{(k)}(\vec{r}) = $$
$$= \sum_{j=k-N}^{k+N} e_{q_0}^{(j)} \left[ \vec{E}_{q_0}^{(k)}(\vec{r}) \delta_{j,k} + \sum_{q \neq q_0} \frac{\omega_q^{(k)2}}{\omega_q^{(k)2} - \omega^2} \alpha_{q_0,q}^{(k,j)} \vec{E}_q^{(k)}(\vec{r}) \right] = \sum_{j=k-N}^{k+N} e_{q_0}^{(j)} \vec{\beta}^{(k,j)}(\vec{r}) \tag{7}$$

where $\vec{r} \in V^{(k)}$, $V^{(k)}$ is a part of chain space which belongs to the $k$-th resonator.

The set of coupling equations (5) can be considered as a difference equation. This difference equation, which defines the amplitudes of the basic modes $e_{q_0}^{(k)}$, is the main equation of the CCM.

It is reasonable to note that the amplitudes of the basic modes $e_{q_0}^{(k)}$ are non-measured values. Indeed, we can measure the components of electric field in any point, for example, by the nonresonant perturbation method [21,22,23], but we cannot measure the amplitudes $e_q^{(k)}$ and have to use numerical methods for finding them by using the expansion (1). This circumstance creates difficulties in studying the properties of the real slow-wave waveguides, including their tuning [23,24]. The similar situation arises also in other electrodynamic models. For example, the space harmonics in the "waveguide" model of homogeneous periodic waveguides [25] are non-measured values, too.

If we choose the set of points $\vec{r}^{(k)}$ which belongs to the volumes of different resonators (for example, $r = 0$, $z^{(k)} = \sum_{s=-\infty}^{k-1} d_s + d_k/2$), we can consider any component of the vectors $\vec{E}^{(k)}(\vec{r}^{(k)})$ as the sequence of complex numbers each of which are the sum of several values of grid function that is the solution of the known difference equation. Can we construct a new difference equation the solution of which will be the components of the electric field at the given points of different resonators?

## 2 The equation for the electric field in the infinity chain of resonators

We can rewrite equations (5) and (7) as

$$\sum_{s=-N}^{N} e_{q_0}^{(k+s)} \frac{\left( \alpha_{q_0,q_0}^{(k,k+s)} - Z_{q_0}^{(k)} \delta_{s,0} \right)}{\left( \alpha_{q_0,q_0}^{(k,k)} - Z_{q_0}^{(k)} \right)} = \sum_{s=-N}^{N} e_{q_0}^{(k+s)} \overline{\alpha}^{(k,k+s)} = 0. \tag{8}$$

$$\sum_{s=-N}^{N} e_{q_0}^{(k+s)} \beta_g^{(k,k+s)}(\vec{r}^{(k)}) = E_g^{(k)}(\vec{r}^{(k)}), \tag{9}$$

where $E_g^{(k)}(\vec{r}^{(k)})$ are the $g$-component of electric field at $\vec{r} = \vec{r}^{(k)}$.

Using the results of Appendix 2, we get the difference equation for $E_g^{(k)}(\vec{r}^{(k)})$

$$\sum_{s=-N}^{N} h_{k,s} E_g^{(k+s)}(\vec{r}^{(k+s)}) = 0. \tag{10}$$

Here $H_k^{(1)} = \left(h_{k,-N},...,h_{k,-1},h_{k,1},...,h_{k,N}\right)^T$ are the solutions of systems of linear equations

$$W_k^{(1)} H_k^{(1)} + W_k^{(2)} H_k^{(2)} = -\mathrm{B}_k. \tag{11}$$

Matrices $W_k^{(1)}, W_k^{(2)}, \mathrm{B}_k, H_k^{(2)}$ are defined in Appendix 2, $h_{k,0} = 1$.

It should be noted that the coefficients $h_{k,s}$ in the equation (10) also depend on the set of points $\vec{r}^{(k-N)}, \vec{r}^{(k-N+1)},..., \vec{r}^{(k)},...\vec{r}^{(k+N-1)}, \vec{r}^{(k-N)}$

The solution of the difference equation (10) gives magnitudes of electric field component $E_g^{(k)}(\vec{r}^{(k)})$ in a fixed set of points $\vec{r}^{(k)}$.

## 3 Solutions of the difference equation for the electric field

We have added to the CCM codes [15,16] procedures for finding the solutions of systems of linear equations (11) and the solutions of the difference equation for $E_g^{(k)}(\vec{r}^{(k)})$ (10).

The procedures of calculation of the coefficients $\alpha_{q_0,q_0}^{(k,k+s)}$ and its dependencies on parameters of the DLW were presented earlier [15,16]. Let's consider the dependencies of the coefficients $\beta_g^{(k,k+s)}(\vec{r}^{(k)})$. Below as the set $\vec{r}^{(k)}$, we will consider the cavity centers for which $r = 0$, $z^{(k)} = \sum_{s=-\infty}^{k-1} d_s + d_k/2$. As the basic mode we choose the mode $q_0 = (0,1,0)$. We also restrict our consideration by the case $E_g^{(k)}(\vec{r}^{(k)}) = E_z^{(k)}(\vec{r}^{(k)})$. For homogeneous DLWs we define that $E_{z,q_0}^{(k)}(\vec{r}^{(k)}) = 1$.

For homogeneous DLWs with the same phase shift per cell $\varphi = 2\pi/3$ but different aperture sizes $\beta_g^{(k,k+s)}$ are presented in Table 1

| Table 1 |||| 
|---|---|---|---|
| $\beta_g^{(k,k+s)}$ for homogeneous DLWs with the phase shift per cell $\varphi = 2\pi/3$, $f = 2856$ MHz, $d = 3.0989$ cm, $t = 0.4$ cm, $r_t = 0$ ||||
| s | $a = 1.6$cm $b = 4.2343$cm | $a = 1.3$cm $b = 4.1406$cm | $a = 1$cm $b = 4.0754$cm |
| -3 | 1.67E-06 | 1.98E-07 | 9.06E-09 |
| -2 | -2.94E-04 | -8.78E-05 | -1.48E-05 |
| -1 | 6.32E-02 | 4.52E-02 | 2.70E-02 |
| 0 | 8.16E-01 | 8.45E-01 | 8.86E-01 |
| 1 | 6.32E-02 | 4.52E-02 | 2.70E-02 |
| 2 | -2.94E-04 | -8.78E-05 | -1.48E-05 |
| 3 | 1.67E-06 | 1.98E-07 | 9.06E-09 |

First of all, we have to note that the RF coupling leads to significantly changes in field distribution in compare with the one that gives formula (2). Indeed, $\beta_z^{(k,k)}$ significantly differs from 1 even for small apertures ($a = 1$cm). This circumstance is especially important for the case of inhomogeneous DLWs.

As follows from Appendix 2, in the case of homogeneous DLWs the coefficients of the difference equation (10) $h_{k,s}$ ($s \neq 0$, $h_{k,0} = 1$) must to be equal the coefficients of the difference equation (8) $h_{k,s} = \overline{\alpha}_{q_0,q_0}^{(k,k+s)}$. Results of calculations that are presented in Table 2 confirm this fact.

For DLWs with varying geometrical sizes the difference between the coefficients $h_{k,s}$ and $\overline{\alpha}_{q_0,q_0}^{(k,k+s)}$ are determined by the values of parameter gradients

Table 2
Homogeneous DLWs with the phase shift per cell $\varphi = 2\pi/3$, $f = 2856$ MHz, $d = 3.0989$ cm, $t = 0.4$ cm, $r_t = 0$, $a = 1.6$ cm, $b = 4.2343$ cm

| s | $h_{k,s}$ | $\overline{\alpha}_{q_0,q_0}^{(k,k+s)}$ |
|---|---|---|
| -3 | 3.06E-05 | 3.06E-05 |
| -2 | -5.44E-03 | -5.44E-03 |
| -1 | 1.01E-00 | 1.01E-00 |
| 0 | 1.00E-00 | 1.00E-00 |
| 1 | 1.01E-00 | 1.01E-00 |
| 2 | -5.44E-03 | -5.44E-03 |
| 3 | 3.06E-05 | 3.06E-05 |

Let consider the more complicated DLW - the biperiodic DLW with such geometric sizes:

cells A: $d = 3.9484$ cm, $t = 0.4$ cm, $r_t = 0$, $a = 1.5$ cm, $b = 4.1329$ cm;

cells B: $d = 0.5$ cm, $t = 0.4$ cm, $r_t = 0$, $a = 1.5$ cm, $b = 4.4712$ cm.

For $N = 3$ coefficients $\beta_g^{(k,k+s)}$, $\overline{\alpha}_{q_0,q_0}^{(k,k+s)}$, $h_{k,s}$ ($f = 2856$ MHz) are given in Table 3. We should like to note that the value of electric field in the cell of B type are mostly determined by the three amplitudes of $E_{010}$ modes in the adjacent cells, while in cells of A type - mostly by the one amplitude (compare the second and the fifth columns)

Table 3

| | cells A: | | | cells B | | |
|---|---|---|---|---|---|---|
| s | $\beta_g^{(k,k+s)}$ | $\overline{\alpha}_{q_0,q_0}^{(k,k+s)}$ | $h_{k,s}$ | $\beta_g^{(k,k+s)}$ | $\overline{\alpha}_{q_0,q_0}^{(k,k+s)}$ | $h_{k,s}$ |
| -3 | -7.39E-05 | -5.23E+01 | 3.14E-02 | -6.37E-04 | -7.57E+02 | 1.15E-01 |
| -2 | 1.04E-02 | 7.33E+03 | -2.77E-00 | -1.86E-03 | -2,12E+03 | 9.33E-01 |
| -1 | 2.12E-02 | 1.80E+04 | -2.86E+01 | 2.66E-01 | 2.97 E+05 | -1.70E+01 |
| 0 | 8.84E-01 | 1.00E-00 | 1.00E-00 | 2.48E-01 | 1.00E-00 | 1.00E-00 |
| 1 | 2.12E-02 | 1.80E+04 | -2.86E+01 | 2.66E-01 | 2.97 E+05 | -1.70E+01 |
| 2 | 1.04E-02 | 7.33E+03 | -2.77E-00 | -1.86E-03 | -2,12E+03 | 9.33E-01 |
| 3 | -7.39E-05 | -5.23E+01 | 3.14E-02 | -6.37E-04 | -7.57E+02 | 1.15E-01 |

It can be seen that the coefficients $\overline{\alpha}_{q_0,q_0}^{(k,k+s)}$ and $h_{k,s}$ are quite different. But these different coefficients give the same dispersive characteristic. Indeed, the characteristic equation ($E_g^{(k+1)} = \rho E_g^{(k)}$) for the difference equation of $\alpha$ – type (8) or $h$ – type (10) has the same form ($N = 3$)

$$\rho^6 + q_1\rho^5 + q_2\rho^4 + q_3\rho^3 + q_2\rho^2 + q_1\rho + 1 = 0. \qquad (12)$$

Results of calculations of the coefficients $q_1, q_2, q_3$ for the difference equations of $\alpha$ – type (with using $\overline{\alpha}_{q_0,q_0}^{(k,k+s)}$) and $h$ – type (with using $h_{k,s}$) show that their difference is small and the characteristic multipliers for these two types of equations coincide. For example, the suitable multipliers for several frequencies are given in Table 4.

Analyzing the obtained results we can conclude that the new difference equation for the electric field values $E_g^{(k)}(\vec{r}^{(k)})$ in the case of homogeneous periodic waveguide gives the same

(up to calculation errors) dispersive characteristic as the difference equation for the amplitudes of $E_{010}$ modes.

Table 4
The solutions of the equation (12)

| $f$ (GHz) | $\rho_\alpha$ | $\rho_h$ |
|---|---|---|
| $2.856^2$ | (-1.00098,0.00000) | (-1.00098,0.00000) |
|  | (-0.99902,0.00000) | (-0.99902,0.00000) |
| 2.857 | (-0.99891,4.65892E-02) | (-0.99891,4.65842E-02) |
|  | (-0.99891,-4.65892E-02) | (-0.99891,-4.65842E-02) |
| 2.86 | (-0.98902,0.14775) | (-0.98902,0.14775) |
|  | (-0.98902,-0.14775) | (-0.98902,-0.14775) |

Consider the accuracy of the description of the distribution of the electric field based on the difference equation for the electric field values $E_g^{(k)}(\vec{r}^{(k)})$ (10).

On the base of the coupling equations (5) or (10) we can study the processes of wave propagation in inhomogeneous segments of DLWs [15,16]. It can be done if we suppose that before and after the inhomogeneous segment of DLW there are the homogeneous segments of DLW (input and output waveguides). In homogeneous segments at sufficient distance from the connection interfaces (when all evanescent waves decay) we can search the solution of the difference equation in the form

$$f^{(k)} = \begin{cases} \exp\{i\varphi_{1,0}(k-k_1)\} + R_f \exp\{-i\varphi_{1,0}(k-k_1)\}, & k < k_1 \\ T_f \exp\{i\varphi_{2,0}(k-k_2)\} & k > k_2 \end{cases} \quad (1.13)$$

where $R_f, T_f$ are the reflection and transmission coefficients.

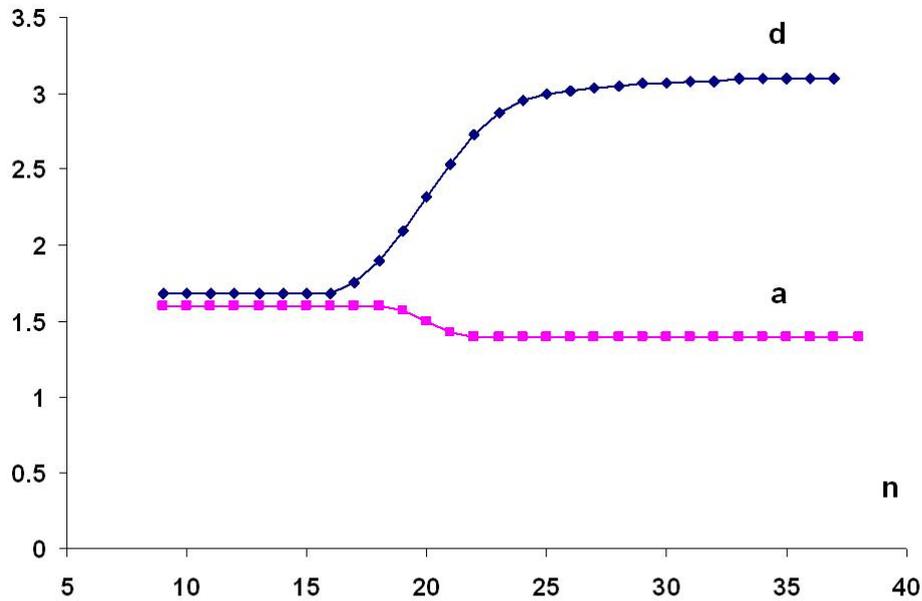

Fig. 1

Injector sections are usually the DLW with geometrical sizes that provide changing along structure not only phase velocity, but the amplitude of accelerating field too. We chose geometrical sizes close to the parameters of the injector that described in [26]. Dependences of the radii of apertures and the cavity length on the cell number are presented in Fig. 1 ($t$=0.4 cm, $r_t$=0).

---
[2] The SUPERFISH gives two frequencies for $\pi$-mode: $f = 2856.38$ MHz, $f = 2856.09$ MHz

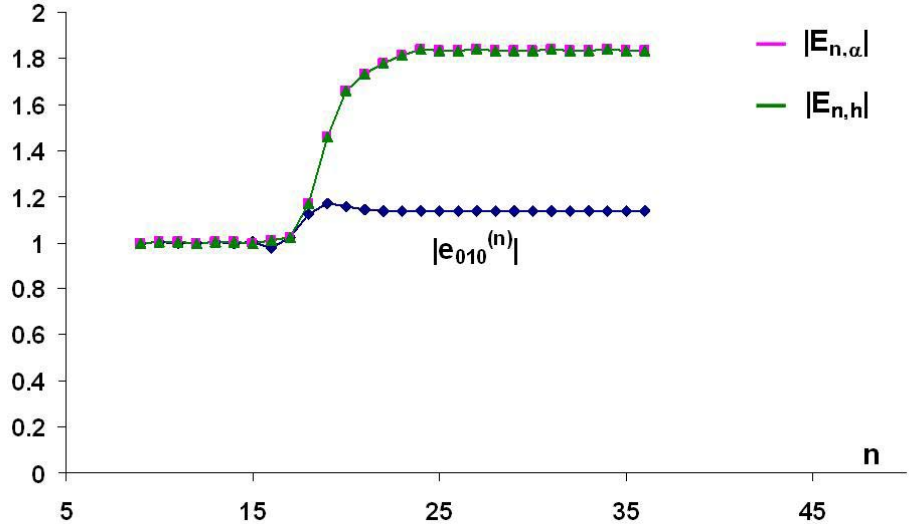

**Fig. 2**

Absolute values of longitudinal electric field in the middle of the cells, that were calculated on the basis of the solution of difference equation (5) with using the sum (7) ($|E_{n,\alpha}|$) and on the basis of difference equation (10) ($|E_{n,h}|$), are presented in Fig. 2. We see that the absolute values of longitudinal electric field for two methods coincide with good accuracy. The same is true for phases. Therefore, the difference equation (10) can be used for direct calculation of the electric field distribution. The difference equation for the magnetic field can be obtained by changing the coefficients $\vec{\beta}^{(k,j)}(\vec{r})$ in the sum (7).

### Conclusions

We constructed a procedure for getting a difference equation, the solution of which is the components of the electric field at the given points of the resonator chain.

Let's follow again the main stages of getting this equation. We started from decomposing the considered region of space into simple regions, found solutions of the Maxwell equations in the form of an infinite series with unknown coefficients that partially satisfy the boundary conditions, used additional boundary conditions and constructed a difference equation for some unknown coefficients. Then we constructed a difference equation for electric field components in some points. To find the values of the electric field, we must solve this equation already without any boundary conditions at the boundaries of the considered region. We started with the wave equation with boundary conditions and obtained the difference equation without boundary conditions. Boundary conditions were included into the coefficients of the difference equation for the electric field. As the obtained equation connects the field values in different points, the proposed procedure represents a nonlocal model for field description.

The proposed technique has several drawbacks. One of them is the necessity to define initial values of the electric field at the given points of the considered region. To eliminate it the procedure must be generalized to a finite number of resonators. The second one is the necessity to work with ill-conditioned matrices. In the Coupled Cavity Model it is assumed that the coupling coefficients decrease rapidly with increasing distance from the given resonator $\bar{\alpha}_{q_0,q_0}^{(k,k+s)} \xrightarrow[s\to\infty]{} 0$. Therefore, in the proposed technique we have deal with numbers that vary greatly in magnitude. Presented above results show that we can, at least in some cases, get fairly accurate values of field parameters. This problem requires more study.

The advantages of working with field values are obvious. We can perform more accurate synthesis of the electromagnetic field distribution in a chain of coupled resonators [27,28,29] and their tuning [23,24]. We can also study the characteristics of the wider class of inhomogeneous waveguides by decomposing them into the chain of coupled resonators [30].

The developed computer codes can be applied for cylindrical geometries for which there are analytical expressions of the eigen functions. The area of using the proposed approach is more wider as there is the general method that gives possibility to obtain the coupling coefficients for arbitrary chain of resonators without using the great number of eigen functions [31].

## Appendix 1

Consider an overdetermined linear system of equations

$$\sum_{s=1}^{M} \gamma_{n,s} y_s = f_n, \quad 1 \leq n \leq M+1. \tag{14}$$

We shall suppose that all equations are independent and such system

$$\sum_{s=1}^{M} \gamma_{n,s} y_s = f_n, \quad 1 \leq n \leq M \tag{15}$$

is consistent. We introduce the set of fundamental solutions

$$\sum_{s=1}^{M} \gamma_{n,s} y_{s,k} = \delta_{n,k}, \quad 1 \leq k \leq M, 1 \leq n \leq M, \tag{16}$$

or in matrix form

$$\Gamma Y = I, \tag{17}$$

where $\Gamma$ and $Y$ are the matrices of $M \times M$ size, $(\Gamma)_{n,s} = \gamma_{n,s}$, $(Y)_{n,s} = y_{n,s}$, $1 \leq k \leq M, 1 \leq n \leq M$. The solution of the system (15) can be represented in the form

$$y_s = \sum_{k=1}^{M} f_k y_{s,k}, \quad 1 \leq s \leq M. \tag{18}$$

Substituting this expression in the last equation ($n = M+1$) of the system (14), we obtain a relation to which the right-hand sides of the system (14) must satisfy

$$\sum_{n=1}^{M+1} h_n f_n = 0, \tag{19}$$

where $h_{M+1} = -1$ and

$$h_n = \sum_{s=1}^{M} y_{s,n} \gamma_{M+1,s}, \quad 1 \leq n \leq M. \tag{20}$$

Using the above approach, for calculation the coefficients $h_n$ we have to solve $M$ linear systems of equations (16). There is another method of calculation of these coefficients. From (17) it follows

$$Y^T \Gamma^T = I \rightarrow \sum_{s=1}^{M} \gamma_{s,n} y_{k,s} = \delta_{k,n} \tag{21}$$

Consider $M$ linear forms

$$\sum_{n=1}^{M} \gamma_{n,s} h_n, \quad 1 \leq s \leq M. \tag{22}$$

Substituting (20) in (22), we get

$$\sum_{n=1}^{M} \gamma_{n,s} h_n = \sum_{n=1}^{M} \sum_{k=1}^{M} \gamma_{M+1,k} \gamma_{n,s} y_{k,n} = \sum_{k=1}^{M} \gamma_{M+1,k} \sum_{n=1}^{M} \gamma_{n,s} y_{k,n}. \tag{23}$$

Using (21), we obtain

$$\sum_{n=1}^{M} \gamma_{n,s} h_n = \gamma_{M+1,s}, \quad 1 \leq s \leq M, \tag{24}$$

or in matrix form

$$\Gamma^T H = F, \tag{25}$$

where $H = (h_1, ..., h_M)^T$, $F = (\gamma_{M+1,1}, ..., \gamma_{M+1,M})^T$.

This is the system of linear equation solution whose solutions are the sought coefficients $h_n$, $1 \le n \le M$.

It is followed from (19) that one element of the sequence $\{h\}$ is the arbitrary one. The system (24) corresponds the case when $h_{M+1} = -1$.

If we want to consider another element $h_K$ of the sequence $\{h\}$ as given, we have to change the equation (24)

$$\sum_{n=1, n \ne K}^{M+1} \gamma_{n,s} h_n = -\gamma_{K,s} h_K, \ 1 \le s \le M, \quad (26)$$

## Appendix 2

For infinitely long chain the main set of coupled equations for amplitudes $y_n$ ($-\infty < n < \infty$) we write as

$$\sum_{s=-M}^{M} \alpha_{n,n+s} y_{n+s} = 0. \quad (27)$$

The additional parameters $f_n$ are the sum of the nearest amplitudes

$$\sum_{s=-M}^{M} \beta_{n,n+s} y_{n+s} = f_n. \quad (28)$$

Let's consider the interval $N \le n \le N + N_c$. What value of $N_c$ have we to choose for obtaining the difference equation for the elements of the sequence $f_n$? This can be done if the number of equations $N_{eq} = 2(N_c + 1)$ will be greater by one than the numbers of unknowns $N_y = (N_c + 1) + 2M$. This condition gives

$$N_{eq} = N_y + 1 \ \to N_c = 2M. \quad (29)$$

Indeed, in this case we can find $N_y = 4M + 1$ unknowns from $N_y$ equations. Substitution them into the last equation gives the relation between $N_c + 1 = 2M + 1$ elements of sequence $\{f_n\}$. It can be considered as a difference equation for $f_n$.

It is convenient to consider the interval $N - M \le n \le N + M$. Then the system of equations that can be transformed into the difference equation for the sequence $\{f_n\}$ take the form ($N \to n$)

$$\begin{aligned}
\sum_{s=-M}^{M} \beta_{n-M,n-M+s} y_{n-M+s} &= f_{n-M} \\
&\cdots \\
\sum_{s=-M}^{M} \beta_{n+M,n+M+s} y_{n+M+s} &= f_{n+M} \\
\sum_{s=-M}^{M} \alpha_{n-M,n-M+s} y_{n-M+s} &= 0 \\
&\cdots \\
\sum_{s=-M}^{M} \alpha_{n+M,n+M+s} y_{n+M+s} &= 0
\end{aligned} \quad (30)$$

Using the results of the Appendix 1, we get the difference equation for $f_n$

$$\sum_{s=-M}^{M} h_{n,s} f_{n+s} = 0. \tag{31}$$

Here $h_{n,s}$ ($-M \leq s \leq 3M+1$, $s \neq 0$) are the solutions of such systems of linear equations

$$W_n^{(1)} H_n^{(1)} + W_n^{(2)} H_n^{(2)} = -B_n h_{n,0}, \tag{32}$$

where $H_n^{(1)} = \left(h_{n,-M}, ..., h_{n,-1}, h_{n,1}, ..., h_M\right)^T$ is the vector of $2M$ length, $H_n^{(2)} = \left(h_{n,M+1}, ..., h_{n,3M+1}\right)^T$ is the vector of $(2M+1)$ length, $W_n^{(1)}$ is the matrix of $(4M+1, 2M)$ size, $W_n^{(2)}$ is the matrix of $(4M+1, 2M+1)$ size, $B_n = \left(0, ..., 0, \beta_{n,n-M}, ..., \beta_{n,n}, ..., \beta_{n,n+M}, 0, ..., 0\right)^T$. The matrix $W_n^{(1)}$ does not contain the column that equals $B_n$.

$$W_n^{(1)} = \overbrace{\begin{pmatrix} \beta_{n-M, n-2M} & \cdots & 0 \\ \beta_{n-M, n-2M+1} & \cdots & 0 \\ \beta_{n-M, n-2M+2} & \cdots & 0 \\ \cdots & \cdots & \cdots \\ \beta_{n-M, n} & \cdots & \beta_{n+M, n} \\ \cdots & \cdots & \cdots \\ 0 & \cdots & \beta_{n+M, n+2M-2} \\ 0 & \cdots & \beta_{n+M, n+2M-1} \\ 0 & \cdots & \beta_{n+M, n+2M} \end{pmatrix}}^{2M} \tag{33}$$

$$W_n^{(2)} = \overbrace{\begin{pmatrix} \alpha_{n-M, n-2M} & \cdots & 0 \\ \alpha_{n-M, n-2M+1} & \cdots & 0 \\ \alpha_{n-M, n-2M+2} & \cdots & 0 \\ \cdots & \cdots & \cdots \\ \alpha_{n-M, n} & \cdots & \alpha_{n+M, n} \\ \cdots & \cdots & \cdots \\ 0 & \cdots & \alpha_{n+M, n+2M-2} \\ 0 & \cdots & \alpha_{n+M, n+2M-1} \\ 0 & \cdots & \alpha_{n+M, n+2M} \end{pmatrix}}^{2M+1} \tag{34}$$

It must be noted that for getting the $2M$ coefficients of the difference equation (31) we have to solve the system of linear equations (32) that include $(4M+1)$ unknowns.

For the homogeneous chain $\alpha_{n,n+s} = \overline{\alpha}_s$, $\beta_{n,n+s} = \overline{\beta}_s$ ($-M < s \leq 3M+1$) the solution of the system (32) is ($h_{n,0} = \pm \alpha_0$)

$$\begin{aligned} h_{n,n+s} &= \pm \overline{\alpha}_s, \quad -M \leq s \leq M, \\ h_{n,n+s} &= \mp \overline{\beta}_s, \quad M < s \leq 3M+1. \end{aligned} \tag{35}$$

For this chain the equations (31) and (27) coincide.